\documentclass[twocolumn,showpacs,preprintnumbers,amsmath,amssymb]{revtex4}

\usepackage{graphicx}
\usepackage{dcolumn}
\usepackage{bm}
\usepackage{dsfont}
\usepackage{mathrsfs}
\usepackage{amsmath}				
 
\usepackage[bookmarks,
                 bookmarksopen = true,
                 bookmarksnumbered = true,
                 linktocpage,
                 colorlinks = true,
                 linkcolor = blue,
                 urlcolor  = blue,
                 citecolor = blue,
                 anchorcolor = green,
                 hyperindex = true,
                 hyperfigures]
		{hyperref}
 
\usepackage[english]{babel}

\usepackage{color}

\newcommand{\be}{\begin{equation}}
\newcommand{\ee}{\end{equation}}
\newcommand{\bey}{\begin{eqnarray}}
\newcommand{\eey}{\end{eqnarray}}

\begin{document} 
\title{Generalized quark number susceptibilities from fugacity expansion\\
at finite chemical potential for $\bf N_f=2$ Wilson fermions}
\author{
Christof Gattringer, 
Hans-Peter Schadler}
\affiliation{
\vspace{3mm}
Institute of Physics,
University of Graz, 8010 Graz, Austria
}

\date{\today}

\begin{abstract}
\vspace{3mm}

Generalized susceptibilities of the net quark number  have been proposed to be good
probes for the transitions  in the QCD phase diagram and for  the search of a
possible critical end point.  In this article we explore a new strategy for computing 
quark number susceptibilities from lattice QCD via an  expansion in the fugacity 
parameter $e^{\mu\beta}$. All quark number related bulk  observables are particularly
easy to access in this approach and we present results for generalized quark number
susceptibilities up to 4-th order. Ratios of these quantities are studied and compared
with model calculations for the high-  and low temperature regions up to a chemical
potential of $\mu\beta\approx1.0$.

\end{abstract}


\maketitle
 
\section{Introduction}
\label{sec:int}
In recent years heavy ion collision experiments started to directly probe the phase
diagram of QCD. In particular results from RHIC and LHC shed  some light on
properties of the quark gluon plasma and the position of the crossover line. To fully
explore the phase diagram theoretically we need to calculate thermodynamical
quantities not only at finite temperature but also at finite baryon chemical
potential. For finite temperature we can use lattice QCD, the only ab-initio method
available. However, for finite baryon density lattice QCD faces the so-called
complex action problem" (or ''sign problem''). In that case the Boltzmann factor
becomes complex and cannot be interpreted as a probability for importance sampling
and the usual Monte Carlo approach fails. To learn something about the phase diagram
at least for small chemical potential, different methods have been developed and over
the last years results have been obtained using reweighting methods, complex Langevin
techniques as well as Taylor expansion in the chemical potential $\mu$ (see, e.g.,
\cite{review} for recent reviews).  

In this paper we focus on a different expansion technique, the
fugacity expansion. In this approach one expands the partition sum in a Laurent
series in the fugacity parameter $z = e^{\mu\beta}$, where $\beta$ is the inverse
temperature and $\mu$ is the chemical potential. This expansion is a rather natural
choice, since the chemical potential on the lattice may be introduced as a boundary
term  by multiplying the factor  $e^{\mu\beta}$ ($e^{-\mu\beta}$) on the forward
(backward) temporal hopping term of the fermion action that connects the last with the first
timeslice. Furthermore, on a finite lattice the fugacity series is finite, while the Taylor expansion
always gives rise to an infinite series. 

For a first implementation of the fugacity expansion 
in QCD based on the strategy used here, see
\cite{oldfugacity}. Preliminary results from this study and a variant for
staggered fermions are documented in \cite{lattice2014}.

Recent comparisons of different expansion techniques to results from a dual 
variables simulation for a QCD related model have shown that the fugacity expansion may provide better
convergence properties than the usual Taylor expansion \cite{modelfugacity}. 
Here we want to explore the possibilities of the fugacity expansion in full QCD by
calculating bulk observables related to the quark number which are particularly
simple in the fugacity expansion. In particular the quark number density and also
higher generalized susceptibilities are evaluated and are compared to hadron
resonance gas model calculations. The results we present are for ensembles 
of $N_f = 2$ Wilson fermions on $N_s^3\times N_t=8^3\times 4$ and $12^3\times 6$
lattices for temperatures below and above the crossover.

\section{Fugacity expansion for Wilson fermions}\label{sec:fug}

The fugacity expansion is introduced in the following way: The partition function for
$N_f = 2$ mass degenerate quark flavors has the general form 

\begin{equation}
	Z_\mu=\int D[U] \, e^{-S_g(U)} \, \det[D(\mu,U)]^2 \; ,
\end{equation}
where $S_g(U)$ denotes the usual Wilson plaquette action and  $D(\mu,U)$ is the
Wilson Dirac operator with chemical potential $\mu$ in a background gauge
configuration $U$ (from now on we omit the gauge field dependence as argument, 
but when useful display the dependence on the chemical potential $\mu$). The fermion
determinant $\det[D(\mu)]$ can be written as a Laurent series in the fugacity
parameter $z = e^{\mu \beta}$,

\begin{equation}\label{eq:fug1}
	\det[D(\mu)]= \sum_{q=-q_{\text{max}}}^{q_{\text{max}}} e^{\mu \beta q}D^{(q)} \; ,
\end{equation}
where $\beta=1/T = N_t$ (the Boltzmann constant and the lattice spacing are set 
to $1$). As mentioned in the introduction, the Taylor expansion is an infinite series 
also on a finite lattice. In contrast, the Laurent series of the fugacity expansion is finite,
summing over all possible net quark numbers $q$ between $q = -q_{\text{max}}$ and 
$q  = +q_{\text{max}}$ with $q_{\text{max}} = 2 \times 3\times N_s^3$. Below we will show
that it is justified to truncate the fugacity series at values of $q$ which are much smaller
than $q_{\text{max}}$.
 
The expansion coefficients $D^{(q)}$ are referred to as ''canonical determinants'' 
and can be computed as the Fourier moments with respect to imaginary chemical
potential (compare also \cite{oldscale,hasentouss}),
\begin{equation}\label{eq:candet}
	D^{(q)} = \frac{1}{2\pi}\int_{-\pi}^{\pi}d\phi\,e^{-iq\phi} 
	\det[D(\mu \beta = i\phi)]  \; .
\end{equation}
They have the interpretation of projections of the grand canonical 
fermion determinant $\det[D(\mu)]$ to a fixed net quark number $q$, and
also provide the basis for the canonical approach (see, e.g., \cite{canonical}). 

For large values of the net quark number the evaluation of the integral 
in (\ref{eq:candet})
becomes numerically harder (see below) and in actual calculations $q$ 
is limited by $|q| \leq q_{\text{cut}}$, where $q_{\text{cut}}$ is much smaller 
than $q_{\text{max}}$. Usually we can reach 
values up to $q_{\text{cut}}=60$ in the numerical evaluation 
depending on the coupling and the volume of the considered ensemble.  

\begin{figure}[t]
\centering
\includegraphics[width=85mm,clip]{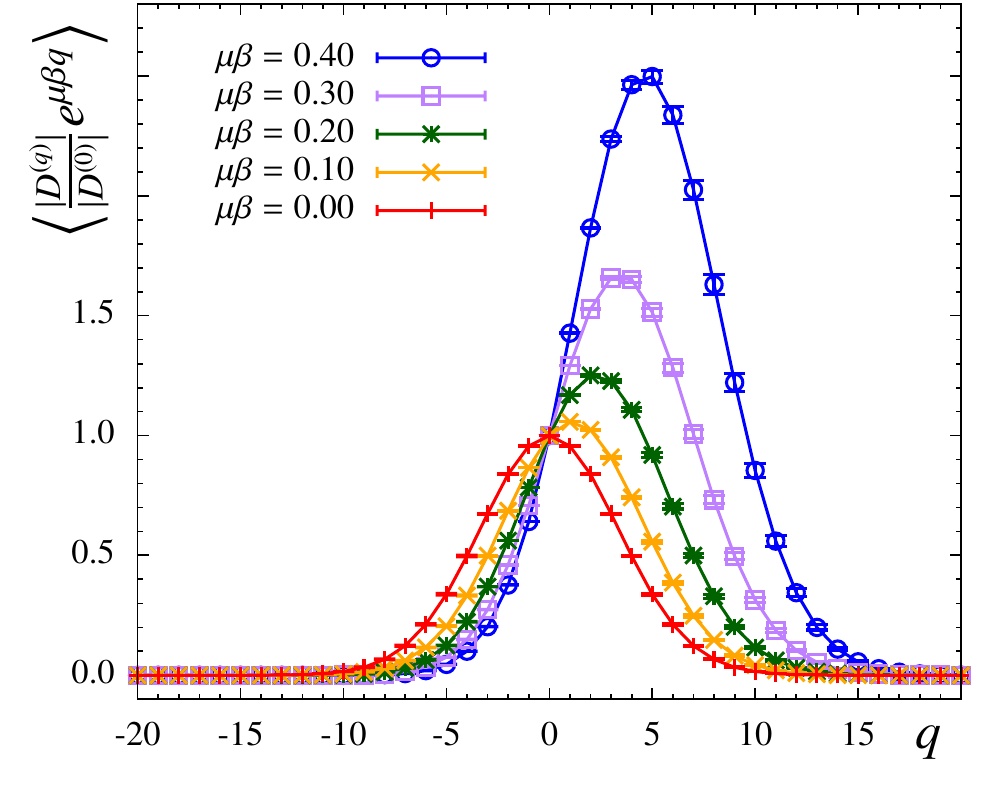}
\includegraphics[width=85mm,clip]{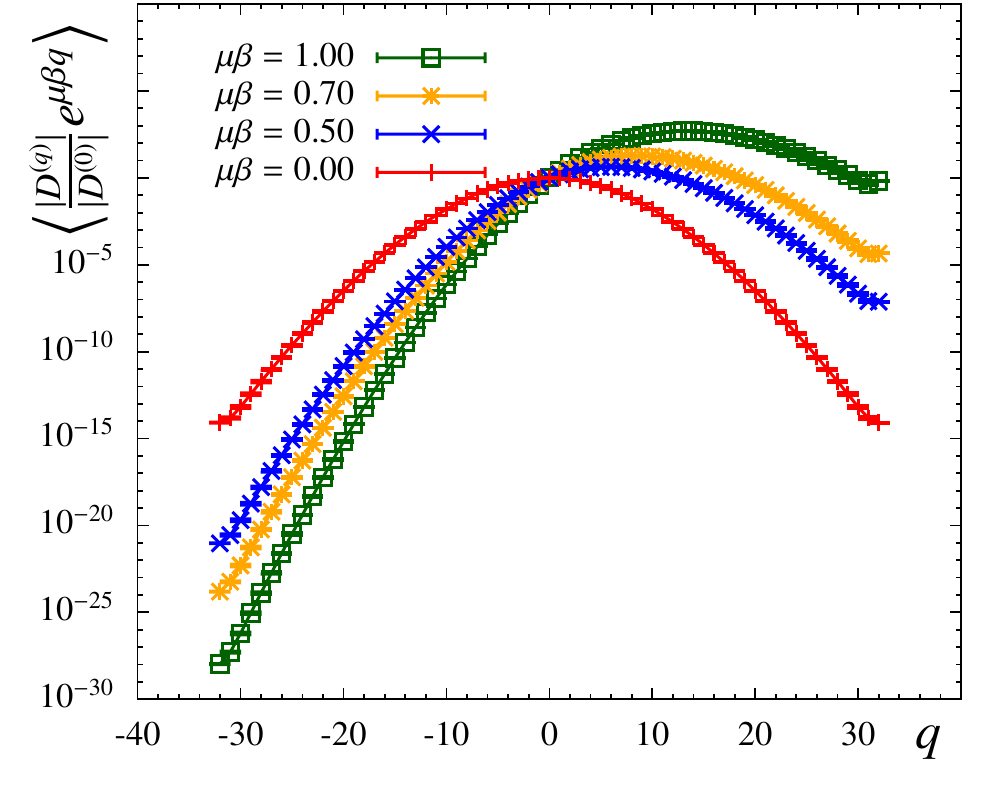}
\caption{Distribution of the absolute values of the canonical determinants normalized by $|D^{(0)}|$ and weighted with the 
fugacity factors as a function of the net quark number 
$q$. The results are from our $12^3\times 6$, $\kappa=0.162$, 
$6/g^2 = 5.40$ ensemble. In the top plot we use a linear scale and values of $\mu \beta$ up to 0.4, while in the bottom plot a log-scale 
is used for values of  $\mu \beta$ up to 1.0.}
\label{Dq}
\end{figure}

Equations (\ref{eq:fug1}) and (\ref{eq:candet}) provide an insight on how the 
chemical potential works physically, which in turn leads to an understanding
of the numerical challenges of the fugacity expansion: 
The key issue is the size distribution of the terms in the sum (\ref{eq:fug1}). 
In Fig.~\ref{Dq} we plot as a function of $q$ the expectation value of the absolute 
value of the summands normalized by $|D^{(0)}|$, i.e., we plot 
$\langle e^{\mu \beta q}  | D^{(q)} | / | D^{(0)}| \rangle$ (for information about the complex phase
of the  $D^{(q)}$ see, e.g., \cite{oldfugacity}).

The $\mu = 0$ data show the distribution of the $|D^{(q)}|$ without the fugacity factors. 
In the plot with the linear scale (top plot in Fig.~\ref{Dq}) they display a 
Gaussian-like distribution centered around $q=0$. In addition to the linear scale, in the
bottom plot we also show the $\langle e^{\mu \beta q}  | D^{(q)} | / | D^{(0)}| \rangle$
using a log scale to better resolve the behavior in the tails of the distribution. 
When turning on the chemical potential the factors $e^{\mu \beta q}$ break the 
symmetry around $q=0$ and shift the distribution of the summands towards larger values of $q$.
This is exactly the behavior that we expect from a chemical potential: The system exhibits an 
average net quark number which is different from zero. Note that the distribution for 
$\mu > 0$ is the result of multiplying the factor $e^{\mu \beta q}$, which exponentially 
increases with $q$, with the Gaussian-type of decay of the $D^{(q)}$. This implies
that the $D^{(q)}$ at large $q$ have to be evaluated very accurately to capture 
the necessary compensation of
the exponential rise from the $e^{\mu \beta q}$. This necessary numerical accuracy 
is the limiting factor for the values of $\mu \beta$ one can reliably reach in the fugacity expansion.

Let us now come to the question of how to express observables related to the
quark number in 
the fugacity series, i.e., in terms of the expansion coefficients $D^{\,(q)}$. We start 
with the partition function where we now express the determinant by the fugacity 
series from Eq.~(\ref{eq:fug1}),

\begin{align}\label{eq:partfun1}
	Z_\mu &= \int D[U] \, e^{-S_g[U]} \, \left( \sum_q e^{\mu\beta q}\,D^{\,(q)} \right)^2  \notag\\
&= \int D[U] \, e^{-S_g[U]} \, \det[D(\mu=0)]^2 \, \times \notag\\
& \qquad \times
\left( \sum_q e^{\mu \beta q}\,\frac{D^{\,(q)}}{\det[D(\mu=0)]} \right)^2   \notag\\
&= \langle (M^{(0)})^2\rangle_0 \, Z_0 \; .
\end{align}
In the last step we have introduced the moments of the canonical determinants

\begin{equation}\label{eq:moments1}
	M^{(n)} = \sum_q e^{\,\mu \beta q}\, q^n \,\frac{D^{\,(q)}}{\det[D(\mu=0)]} \; .
\end{equation}
The expectation values $\langle...\rangle_0$ are evaluated 
on configurations generated with vanishing chemical potential $\mu=0$. This implies that 
the fugacity series has the overlap problem in the same way as all expansion approaches,  
in particular also the Taylor series. 
 
In this study we consider bulk observables related to the quark number which 
are given by derivatives of the partition function with respect to the chemical 
potential $\mu$, i.e., the generalized susceptibilities ($V = N_s^3$):

\begin{equation}
	\chi^q_n \; = \; \frac{1}{V \beta} \frac{\partial^n \ln Z_\mu}{\partial \mu^n} \; .
\end{equation}
Here the label $q$ on the lhs.~is used to distinguish the generalized quark number susceptibilities from 
other commonly used susceptibilities in QCD.
Using Eq.~(\ref{eq:partfun1}) the first derivative, i.e., the quark number density, 
can be easily calculated and is given by

\begin{equation}
	\frac{\chi^q_1}{T^3} = \frac{n_q}{T^3} = 2\;\frac{\;\beta^3}{V}\;
	\frac{\langle M^{(0)} M^{(1)}\rangle_0}{\langle (M^{(0)})^2\rangle_0} \; .
\end{equation}
The result is expressed in terms of the moments defined in 
Eq.~(\ref{eq:moments1}). Another derivative gives the quark number susceptibility

\begin{align}
	\frac{\chi_2^q}{T^2} = 2\frac{\beta^3}{V}&\left[\frac{\langle (M^{(1)})^2\rangle_0 +
	 \langle M^{(0)} M^{(2)}\rangle_0}{\langle (M^{(0)})^2 \rangle_0}\right.\notag\\
	&\;-\left.2\left(\frac{\langle M^{(0)} M^{(1)}\rangle_0}{\langle ( M^{(0)})^2\rangle_0}\right)^2\right] \; .
\end{align}

It is straightforward to express further derivatives via the moments $M^{(n)}$, and for
$\chi_3^q$ and $\chi_4^q$, which we also consider here, we do not display the simple 
but somewhat 
lengthy expressions.  $\chi_3^q$ and $\chi_4^q$ will be particularly useful when we 
study ratios of  generalized susceptibilities which 
have nice properties, not only for numerical tests, but are also convenient for
a comparison with experiment \cite{fluctuations}.

\section{Evaluation and properties of $\bf D^{(q)}$}\label{sec:dqnumcalc}

As we have already discussed, for accessing reasonably large values of 
$\mu\beta$ we need to very precisely calculate the canonical determinants 
$D^{(q)}$ up to a large net quark number $q$. Two factors crucially influence the accuracy
when computing the $D^{(q)}$ using Eq.~(\ref{eq:candet}): The number of values
of $\phi \in [-\pi,\pi]$ has to be sufficiently large when numerically computing the
Fourier integrals and the integrand itself has to be evaluated at high precision.  

For the results presented here we use 256 integration points for $\phi$, but we have also performed
tests for larger and smaller numbers of points. In these tests we did not observe
deviations of the results and we conclude that the integration is stable for
the values of $q$ we take into account here. 

For obtaining the necessary accuracy of the integrand in Eq.~(\ref{eq:candet}),
i.e., for the precise calculation of the canonical
determinant $\det[D(\mu \beta = i\phi)]$, we use an exact evaluation 
with LU factorization. This is of course a sizable numerical
effort and to reduce the computing time and memory requirements we
apply the dimensional reduction given in \cite{dimreduction} (for a different dimensional reduction formula of the Wilson Dirac 
operator determinant see \cite{dimred2}). We here use a slightly modified
form which is more suitable for our calculations (see the appendix for a detailed
derivation). The dimensional reduction can be summarized as follows: The
determinant of the Dirac operator can exactly be rewritten as
\begin{equation}
  \det[D(\mu)]= A\, W(\mu \beta)  \; ,
  \label{dimred1}
\end{equation}
with
\begin{equation}
	W(\mu \beta) = \det[K_0 - e^{\mu \beta} K - e^{-\mu \beta} K^\dagger]  \; ,
	\label{dimred2}
\end{equation}
and a $\mu$-independent factor $A$ which cancels in the calculation of the
observables. $K_0$ and $K$ are two dense matrices depending on the gauge fields, but not
on the chemical potential $\mu$.  We pre-compute them in our code, store them completely
and then use them many times for evaluating $W(\mu \beta)$ with $\mu \beta = i \phi$ at all 
values $\phi$ needed for the numerical integration of  Eq.~(\ref{eq:candet}).
The dimension of the matrices $K_0$ and $K$ is $N_s^3 \times 1 \times 4 \times 3$,
in contrast to the original (sparse) Dirac operator which is $N_s^3 \times N_t \times
4 \times 3$ dimensional. Since the cost of the exact evaluation of the determinant 
is of third order in the dimension of $D$, the exact dimensional reduction 
(\ref{dimred1}), (\ref{dimred2}) speeds up the evaluation by a factor proportional 
to $N_t^3$. With the computer resources available to us this procedure 
allows for a precise and cost efficient evaluation of  $\det[D(\mu)]$ on lattices with
sizes up to $12^3\times 6$.

The results we present in this article are for two different sets of dynamical Wilson
fermion ensembles, $N_s^3\times N_t=8^3\times 4$ with an inverse mass parameter of
$\kappa=0.158$ and $12^3\times 6$ with $\kappa=0.162$. The temperature is driven
by varying the inverse coupling in the range of $6/g^2=5.00$ up to $6/g^2=5.70$.
These parameters correspond to lattice spacings of approximately $a=0.320\,\text{fm}$ down
to $a=0.150\,\text{fm}$ and to pion masses of $M_\pi\leq 960\,\text{MeV}$ (for
$\kappa=0.158$) and $M_\pi\leq 900\,\text{MeV}$ (for $\kappa=0.162$), respectively.
The errors are calculated using the Jackknife method with $300$ configurations per
parameter for the smaller lattices and $50$ to $100$ configurations for the larger
lattices. To study volume effects we use data with different spatial volumes. All
configurations were generated using the MILC collaboration public lattice gauge
theory code \cite{MILC}.

\begin{figure}[t]
\centering
\hspace*{-5mm}
\includegraphics[width=75mm,clip]{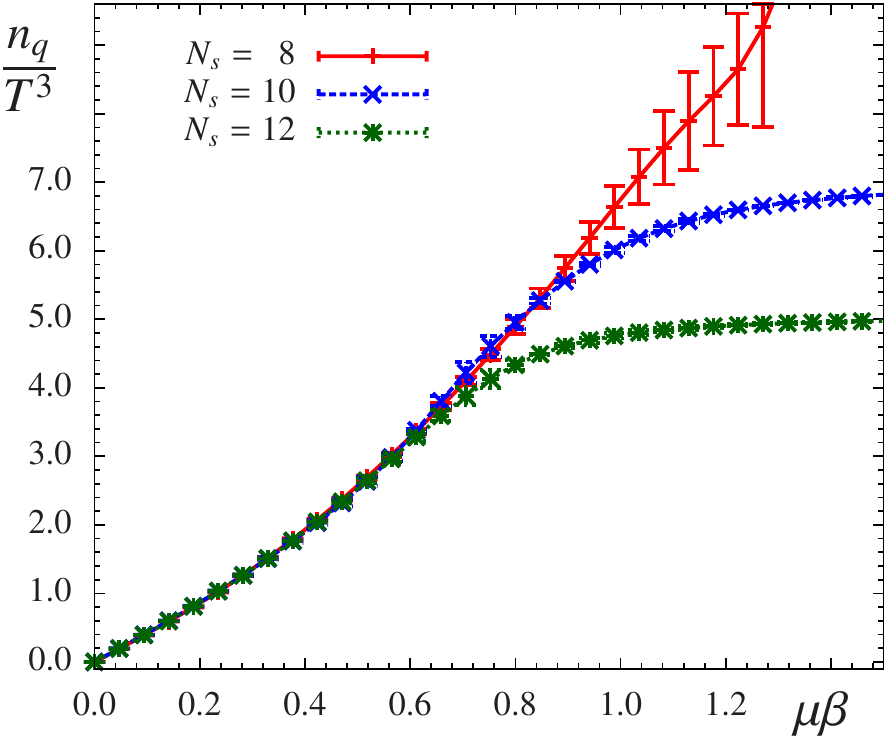}
\caption{Comparison of the quark number density as a function of the 
chemical potential for three spatial volumes with $N_s=8$, $N_s=10$ and 
$N_s=12$. The results are for $N_t=4$, $\kappa=0.158$ and $6/g^2=5.30$.}
\label{volcompnq}
\end{figure}

Let us comment on the volume dependence of the numerical cost: 
The width of the distribution of the $D^{(q)}$ shown in Fig.~\ref{Dq}
is related to the quark number susceptibility and thus is an extensive quantity, 
i.e., it grows with the 3-volume $V$. Thus when increasing the volume,
also $q_{cut}$ has to grow accordingly and the $D^{(q)}$ have to be evaluated for 
larger values of $q$. In turn, when the number
of $D^{(q)}$ that are used is kept fixed while increasing the volume, 
one expects that the fugacity series breaks down 
already at smaller values of $\mu \beta$.

This is illustrated in Fig.~\ref{volcompnq}, where we study the quark number density 
as a function of $\mu \beta$ for three different spatial 
volumes at $\kappa=0.158$ and a coupling near the crossover, $6/g^2=5.30$.
The different spatial lattice extents are $N_s=8$, $N_s=10$ and $N_s=12$. 
Up to a value of $\mu\beta=0.6$ the results from the three volumes 
agree. At this point, however, the density of the largest volume becomes flat 
while for the other volumes it still rises. The statistical errors are still 
small, but nevertheless this signals the breakdown of the expansion at 
this point for the largest volume. When we increase the chemical potential to 
$\mu\beta=0.9$ also the $N_s=10$ result starts to deviate and becomes flat. 
The smallest lattice volume still shows the expected behavior, i.e.,
a rise with chemical potential. A similar behavior is seen in the volume dependence 
of the quark number susceptibility and we conclude that when 
increasing the volume, also $q_{cut}$ has to be increased roughly linearly with 
$N_s^3$ (see also the discussion below).
 
 \begin{figure}[t]
  \centering
  \hspace*{-5mm}
  \includegraphics[width=75mm,clip]{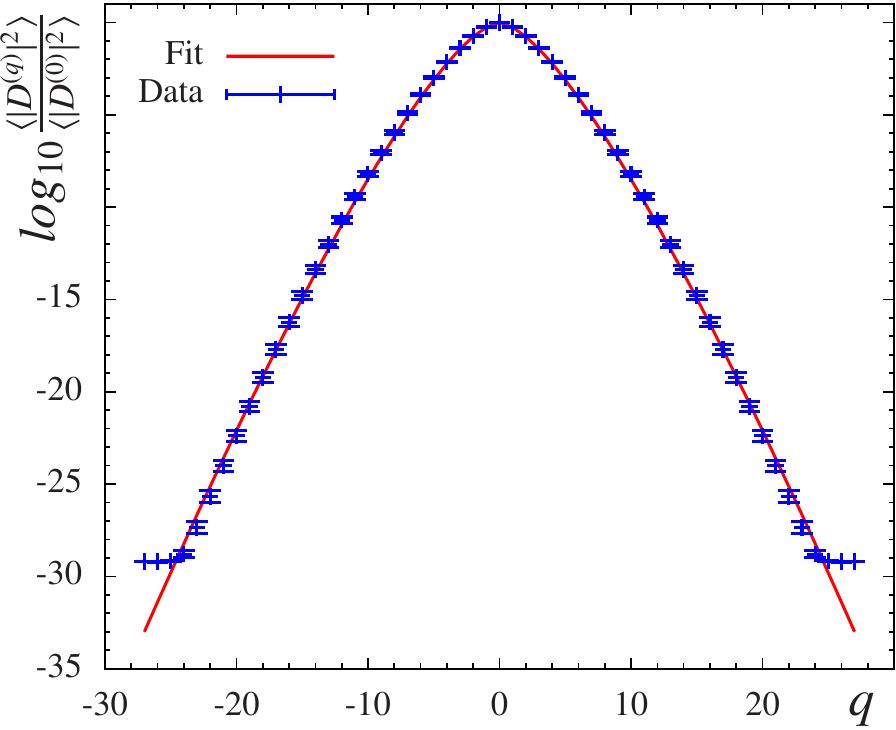}
  \hspace*{-5mm}
  \includegraphics[width=75mm,clip]{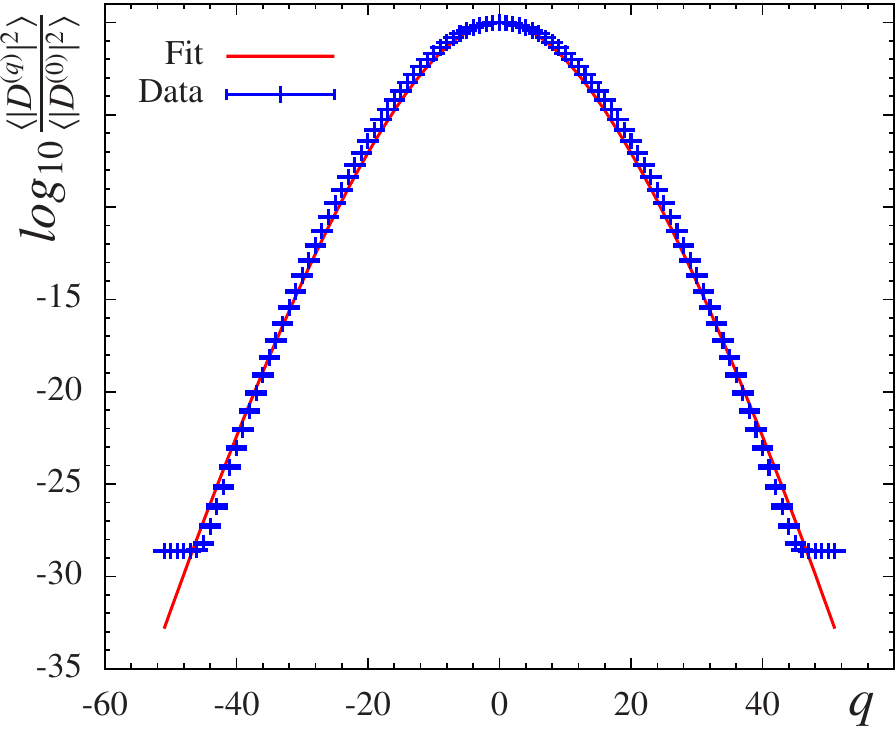}
  \caption{Logarithm of the normalized canonical determinants and a fit of the data to a Skellam distribution (Eq.~(\ref{eq:dqdistri})) are shown. The data is for the $8^3\times 4$, $\kappa=0.158$ ensemble at temperatures $T=144\,\text{MeV}$ (upper plot) and $T=211\,\text{MeV}$ (lower plot).}
\label{Dqfit1}
\end{figure}

\begin{figure*}[t!]
  \centering
  \hspace*{-10mm} \hspace{1cm} $8^3 \times 4, \kappa=0.158$ \hspace{5.5cm} $12^3 \times 6, \kappa=0.162$
  
  \hspace*{-5mm} 
  \includegraphics[width=75mm,clip]{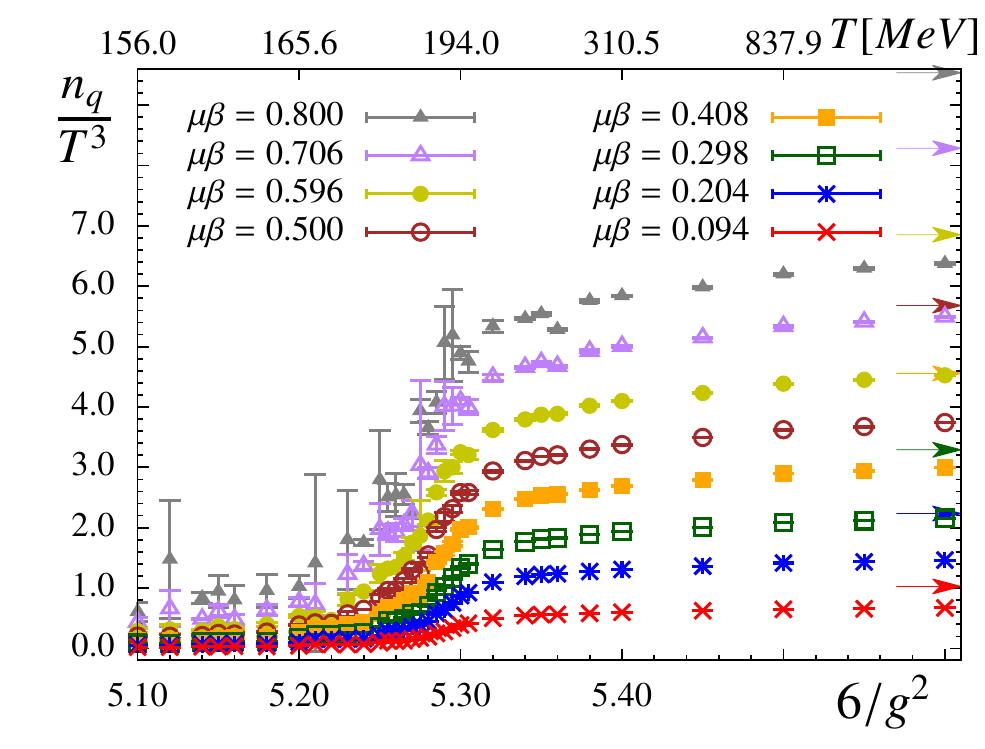}
  \hspace*{5mm}
  \includegraphics[width=75mm,clip]{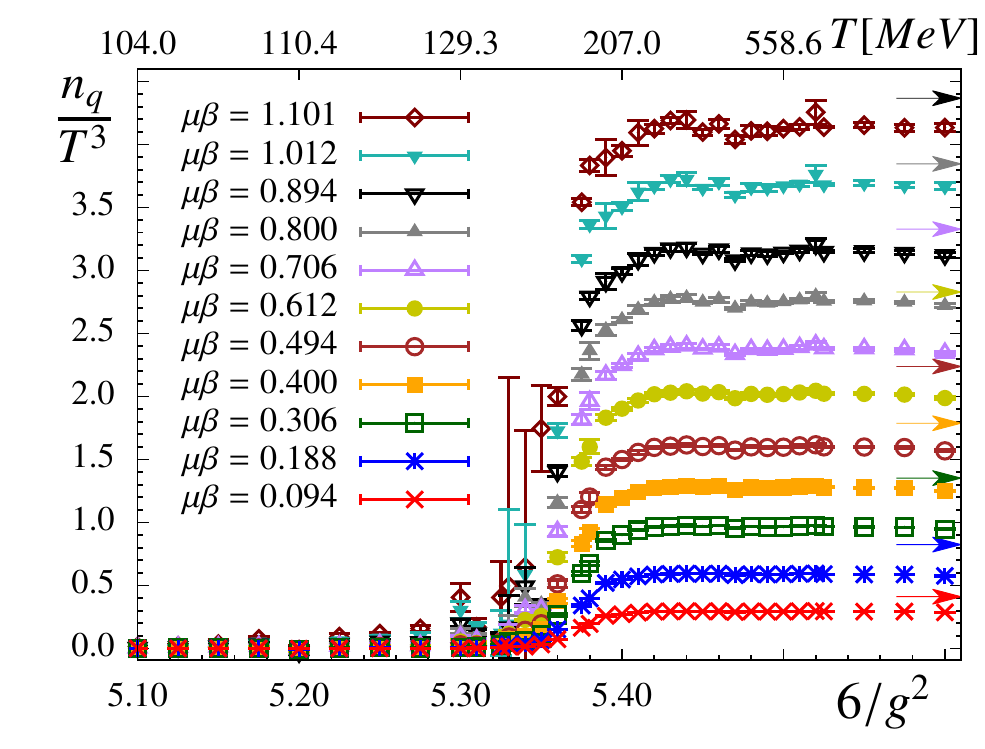}\\
  \hspace*{-5mm}
  \includegraphics[width=75mm,clip]{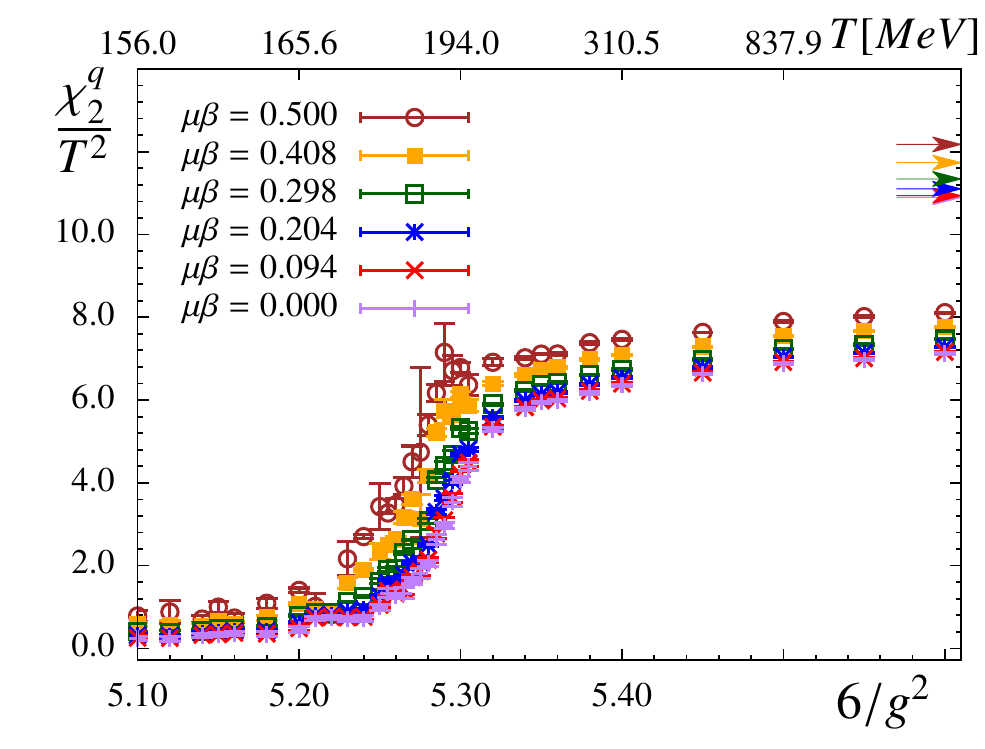}
  \hspace*{5mm}
  \includegraphics[width=75mm,clip]{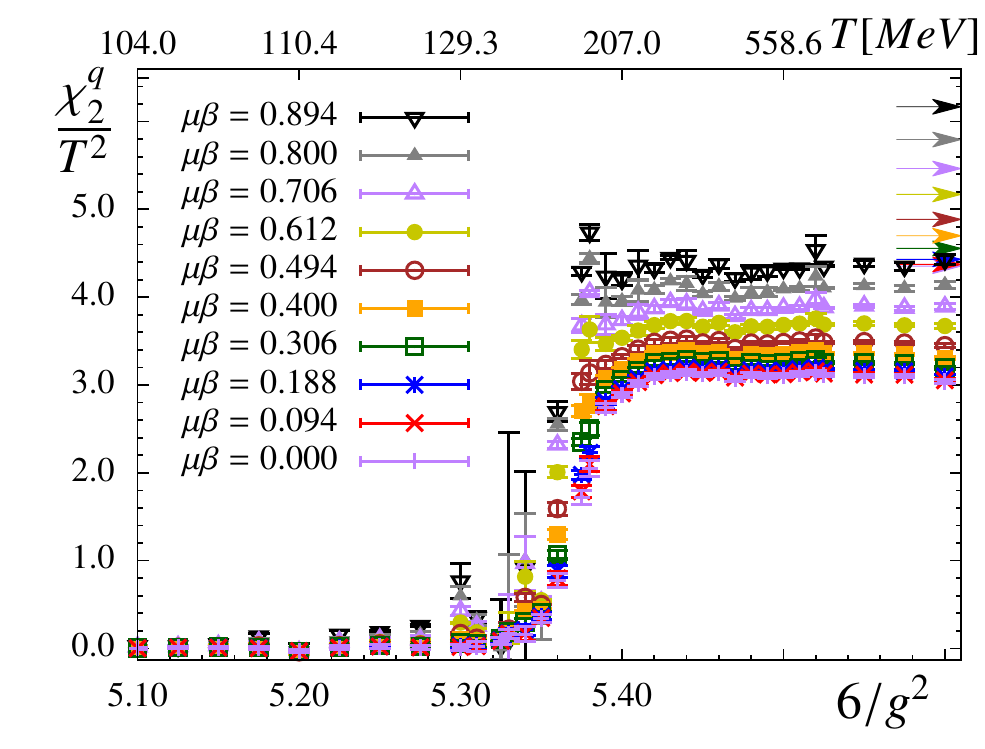}
  \caption{Quark number density $n_q/T^3$ and susceptibility $\chi_2^q/T^2$
  as a function of the inverse coupling (corresponding temperature values on the top scale)
  for the $8^3\times 4$, $\kappa=0.158$ (lhs.) and $12^3\times 6$, $\kappa=0.162$ (rhs.) ensembles.
  The arrows on the rhs.~of the plots mark the Stefan-Boltzmann high temperature limits.}
  \label{susceptibilities}
\end{figure*}

As we have already mentioned above, the distribution of the canonical determinants $D^{(q)}$ 
governs the numerical cost of the fugacity expansion, and various ideas have been followed
to describe the distribution \cite{quarkdistribution,kim}. 
In particular for the distribution of $|D^{(q)}|^2$ one expects a so-called Skellam distribution which, in a normalized form, is given by
			\begin{equation}\label{eq:dqdistri}
				\frac{ \langle |D^{(q)}|^2 \rangle}{\langle |D^{(0)}|^2 \rangle} = \frac{I_q\left(\zeta\right)}{I_0\left(\zeta\right)} \; ,
			\end{equation}
where the $I_q$ are modified Bessel functions. For a qualitative comparison, in Fig.~\ref{Dqfit1} 
we show a fit of this distribution in the parameter $\zeta$ to data from our 
$8^3 \times 4$, $\kappa=0.158$ ensemble. We use two temperatures, $T=144\,\text{MeV}$ below the crossover (upper plot) and 
$T=211\,\text{MeV}$ above the crossover (lower plot). Clearly visible is the strong dependence on the temperature: For low temperatures the width 
of the distribution is small compared to high temperatures where it is nearly twice as wide as in the $T=144\,\text{MeV}$ case.
The flat areas for large values of $|q|$ are numerical artifacts and for the evaluation of the fugacity 
sum one has to stop at smaller $q$. Also 
for the fit to the Skellam distribution these points have been excluded.

From chiral perturbation theory one can derive formulas for various distributions of the canonical determinants, which make explicit also the 
dependence on the temperature and pion mass \cite{kim}. For the distribution considered in Eq.~(\ref{eq:dqdistri}) it gives
the explicit dependence of the argument $\zeta$ on the parameters $M_\pi$, $T$, as well as the spatial volume $V$. In particular 
$\zeta$ is proportional
to the spatial volume, implying that the width of the distribution increases roughly linearly with $V$, as stated above in the discussion of the
numerical cost as a function of the volume. A detailed comparison of the Monte Carlo results 
to the distributions from chiral perturbation theory is presented in \cite{kim}.

\section{Hadron resonance gas}\label{sec:hrg}

The hadron resonance gas (HRG) approach is based on the idea that 
in the confined region only quasi free hadrons appear. In this section we 
briefly collect and summarize the HRG formulas which we need for the comparison 
with our results. The basic ansatz of the HRG is to describe the system as a sum over free fermion 
partition functions for baryons and mesons (including resonances) with their appropriate masses \cite{hrg}. 
Here we are only interested in the part which contributes to the quark number related observables, i.e., the baryonic part. 
This part can be written as
\begin{align}
	\ln Z_B &\approx \sum_i \frac{V d_i T}{\pi^2}\left[m_i^2 K_2\left(m_i\beta\right) \right] \cosh\left(3\mu\beta\right) \notag\\
	&= F(T,m) \cosh\left(3\mu\beta\right) \approx \frac{p}{T^4}\; .
\end{align}
Here $K_2$ denotes a modified Bessel function and the sum runs over all baryons with masses $m_i$. 
The introduction of the chemical potential gives rise 
to the cosh function, which we have separated in the third step from the mass dependent part denoted by $F(T,m)$.

\begin{figure*}[h!t]
  \centering
  
  \hspace*{-10mm} \hspace{1cm} $8^3 \times 4, \kappa=0.158$ \hspace{5.5cm} $12^3 \times 6, \kappa=0.162$
  
  \hspace*{-5mm} 
  \includegraphics[width=75mm,clip]{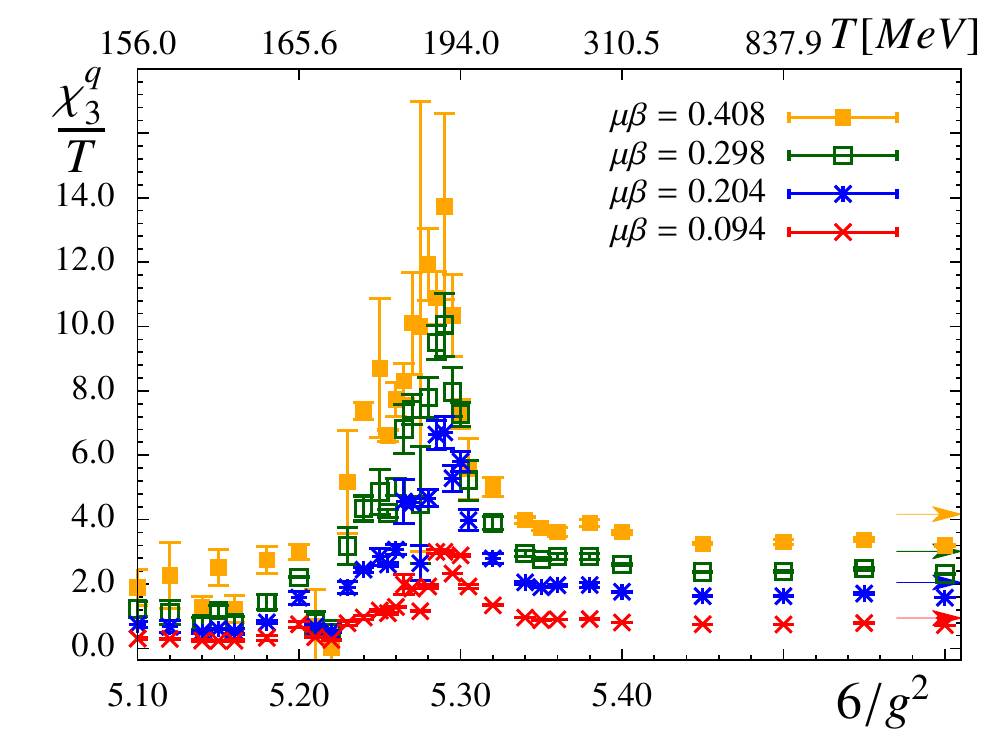}
  \hspace*{5mm}
  \includegraphics[width=75mm,clip]{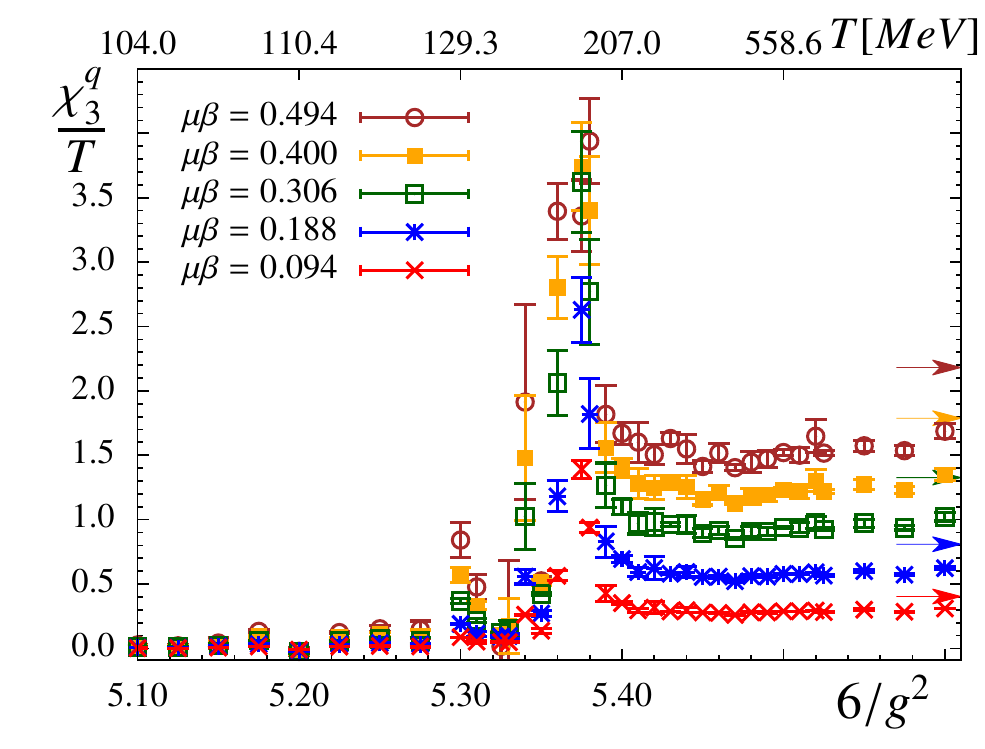}\\
  \hspace*{-5mm}
  \includegraphics[width=75mm,clip]{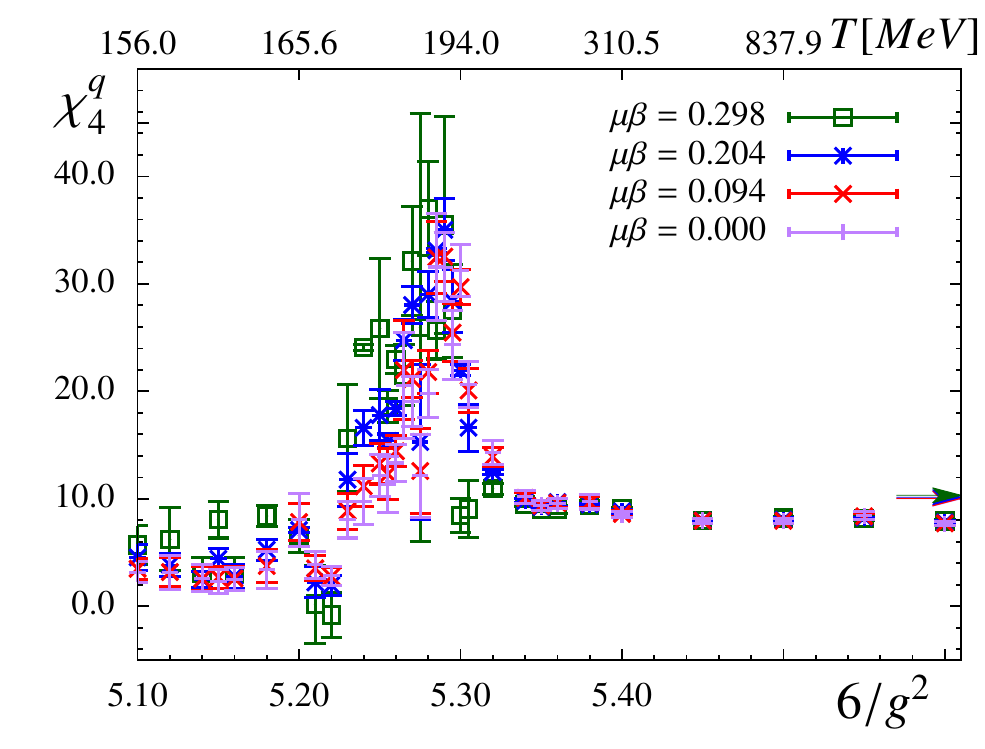}
  \hspace*{5mm}
  \includegraphics[width=75mm,clip]{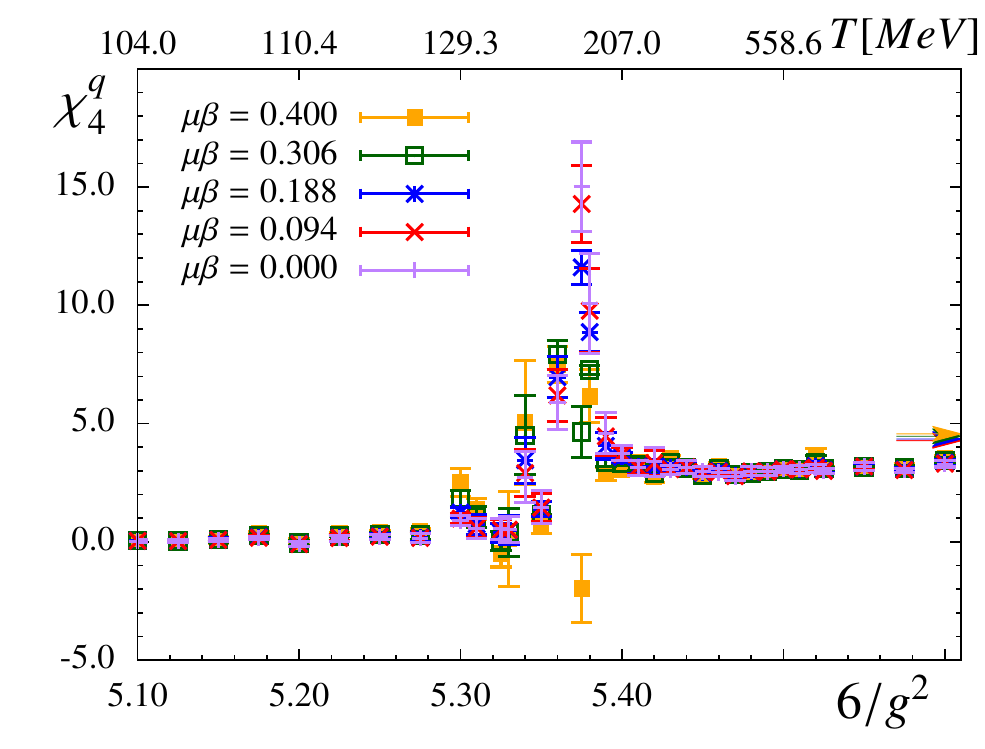}
  \caption{The generalized quark number susceptibilities $\chi_3^q/T$ (top) and $\chi_4$ (bottom) 
  as a function of the inverse coupling (corresponding temperature values on the top scale) 
  for the $8^3\times 4$, $\kappa=0.158$ (lhs.) and $12^3\times 6$, $\kappa=0.162$ (rhs.) ensembles. 
  The arrows on the rhs.~of the plots mark the Stefan-Boltzmann high temperature limits.}
  \label{highersusceptibilities}
\end{figure*}

It is obvious that ratios of derivatives with respect to $\mu$ are independent of this function $F(T,m)$ and therefore independent of the mass 
spectrum. Furthermore, the dependence on the chemical potential will either be via a $\tanh(3\mu\beta)$ or a $\text{coth}(3\mu\beta)$ or, for the 
case of even/even or odd/odd derivative ratios, independent of $\mu$. In particular we will use the following ratios
\begin{align}
	\frac{n_q/T^3}{\chi^q_2/T^2} &= \frac{1}{3}\tanh(3\mu\beta) \; ,\\
	\frac{\chi^q_3/T}{\chi^q_2/T^2} &= 3\,\text{coth}(3\mu\beta)\; ,
\end{align}
and
\begin{equation}
	\frac{\chi^q_4}{\chi^q_2/T^2} = 9\; ,
\end{equation}
for comparison with the observables obtained using the fugacity expansion.

\section{Numerical results}\label{sec:results}

We now come to the presentation of the results for the generalized susceptibilities and their ratios (for a strategy of comparing the
lattice data to experiments see \cite{nana}). We compare the results
for our two lattices, $8^3 \times 4$ and $12^3 \times 6$, with parameters as specified in Section III. Before showing the new results
we remark that the $8^3\times 4$ data have parameters similar to the preliminary calculations with the fugacity expansion
discussed in \cite{oldfugacity}. The new results 
agree with those from \cite{oldfugacity} within error bars, but they have smaller errors due to improvements in the numerics.


\begin{figure*}[t]
  \centering
    
  \hspace*{-10mm} \hspace{1cm} $8^3 \times 4, \kappa=0.158$ \hspace{5.5cm} $12^3 \times 6, \kappa=0.162$

  \hspace*{-5mm}
  \includegraphics[width=75mm,clip]{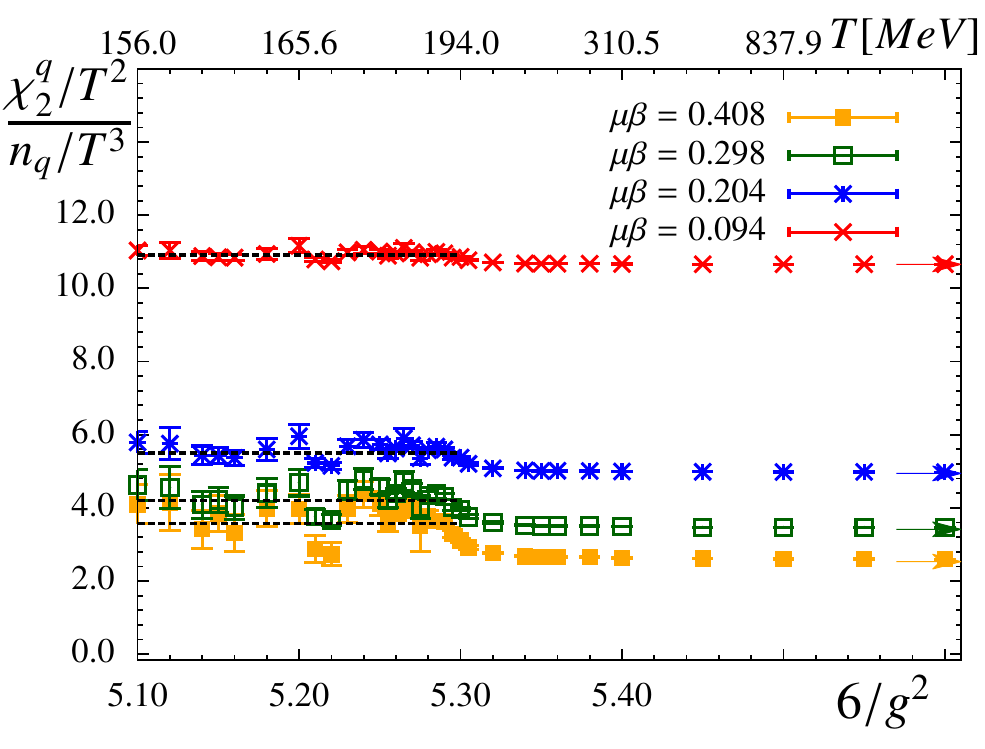}
  \hspace*{5mm}
  \includegraphics[width=75mm,clip]{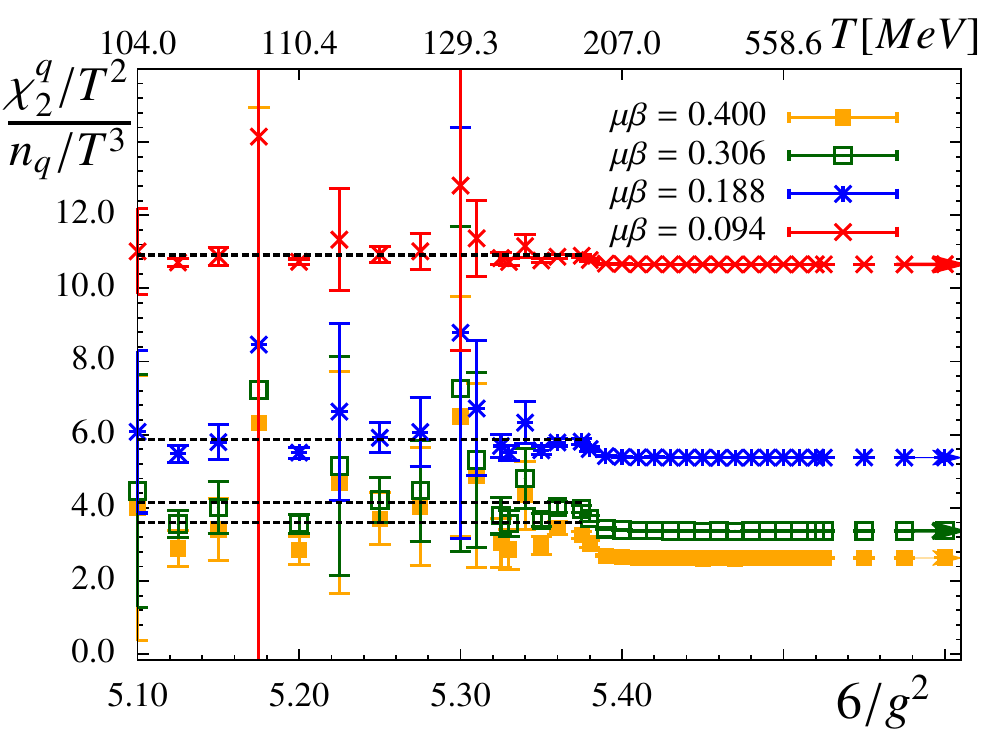}\\
  \hspace*{-5mm}
  \includegraphics[width=75mm,clip]{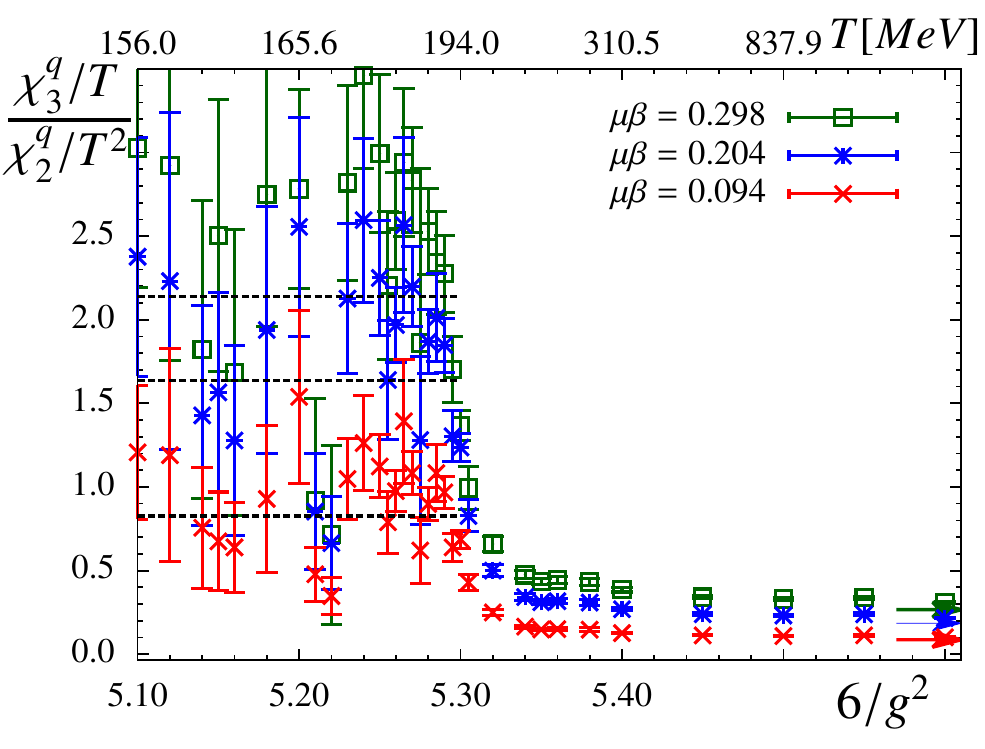}
  \hspace*{5mm}
  \includegraphics[width=75mm,clip]{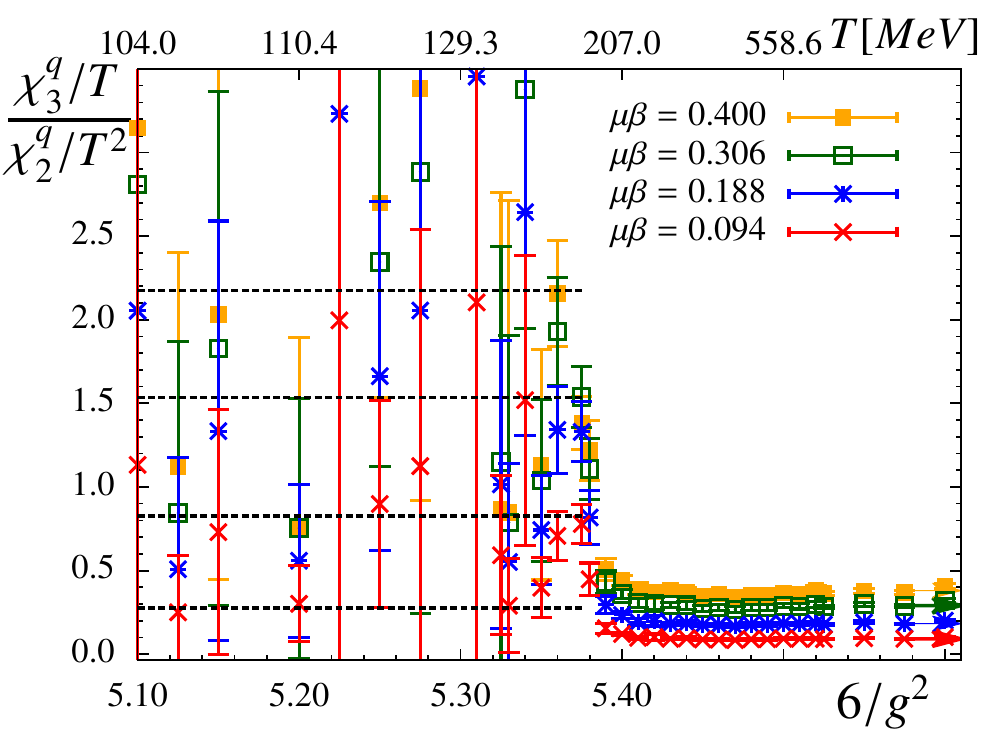}\\
  \hspace*{-5mm}
  \includegraphics[width=75mm,clip]{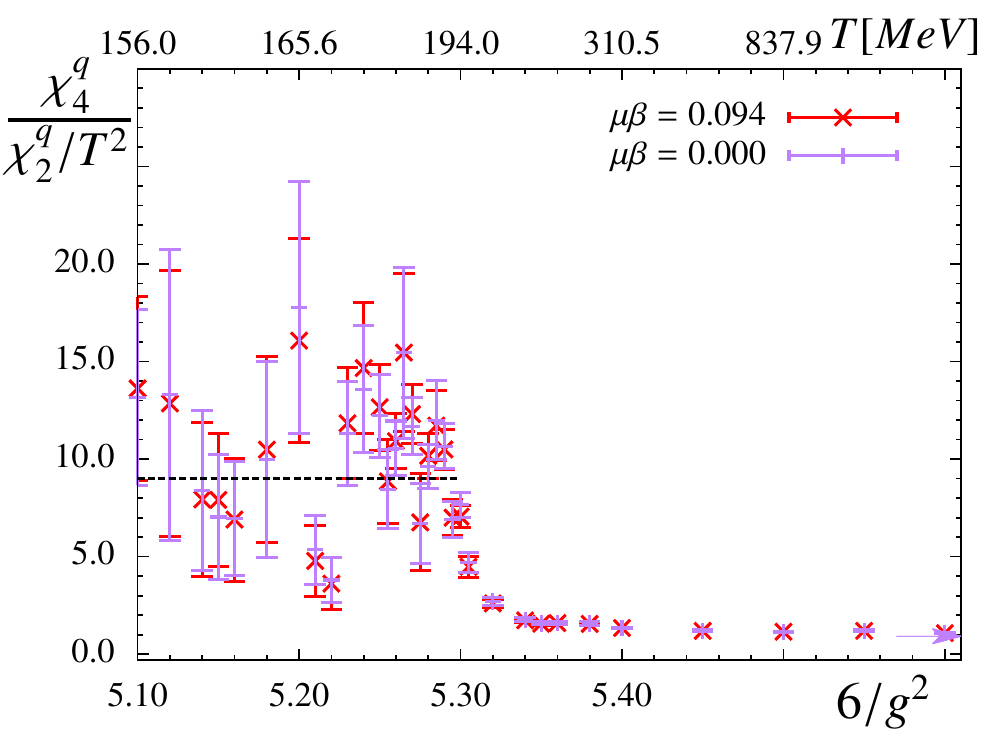}
  \hspace*{5mm}
  \includegraphics[width=75mm,clip]{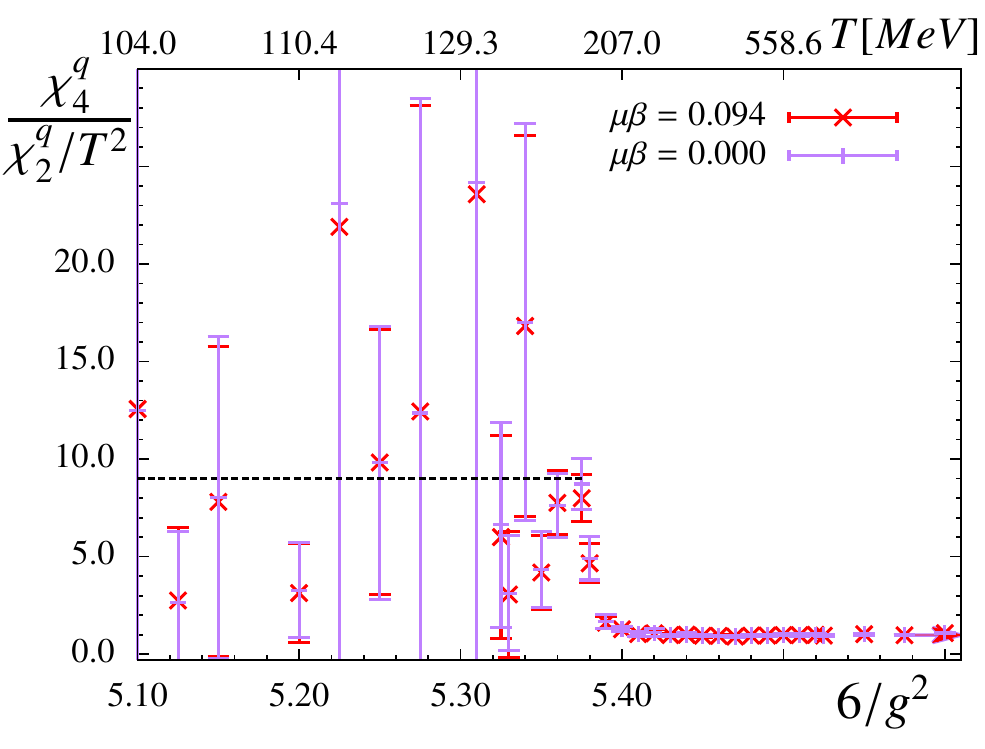}
  \caption{Ratios of the generalized quark number susceptibilities  $(\chi^q_2/T^2)/(n_q/T^3)$, 
  $(\chi^q_3/T)/(\chi^q_2/T^2)$ and $(\chi^q_4)/(\chi^q_2/T^2)$ (top to bottom)
  as a function of the inverse coupling (corresponding temperature values on the top scale)
  for the $8^3\times 4$, $\kappa=0.158$ (lhs.) and $12^3\times 6$, $\kappa=0.162$ (rhs.) ensembles.
  The dashed lines in the confined region are the HRG results and the 
  arrows on the rhs.~of the plots mark the Stefan-Boltzmann high temperature limits.}
  \label{ratios}
\end{figure*}

In Fig.~\ref{susceptibilities} we now show the new results for the quark number density $n_q/T^3$ (upper plots) 
and the quark number susceptibility $\chi_2^q/T^2$ 
(lower plots) as a function of the inverse coupling (corresponding temperature values on the top scale)
for different values 
of the chemical potential $\mu\beta$. The plots on the lhs.~are for $8^3\times 4$, $\kappa=0.158$ and 
$12^3\times 6$, $\kappa=0.162$ is used on the rhs.

For the quark number density $n_q/T^3$, we find a small but non-zero value below the crossover (for $\mu > 0$; for $\mu =0$ one has 
$n_q \equiv 0$ for all temperatures). In the crossover region we observe a more rapid increase with $6/g^2$ (respectively $T$) and above the
crossover again a slow rising. Increasing the chemical potential $\mu$ essentially shifts the whole function for
$n_q/T^3$ towards larger quark numbers. Qualitatively the behavior is the same for both lattice sizes. At very large temperatures one expects a
free gas behavior (Stefan-Boltzmann behavior) and we mark the corresponding values by arrows on the rhs.~of the plots. Obviously the 
temperatures we work at are not yet in the Stefan-Boltzmann region.
 
The second derivative with respect to the chemical potential, i.e., the quark number susceptibility, also shows the expected behavior: Below 
the crossover the curves are flat but non-zero while the crossover is marked by a sudden steep rise of $\chi_2^q/T^2$. 
Above the crossover 
$\chi_2^q/T^2$ develops a plateau-like behavior with only a small slope, again undershooting the Stefan-Boltzmann values  
(marked by the arrows) and approaching them only slowly with temperature.  It is obvious, that with increasing chemical potential the position 
of the steep rise of $\chi_2^q/T^2$ shifts towards smaller $T$. This corresponds to the bending  of the crossover curve in the $\mu$-$T$ 
diagram towards smaller temperatures when increasing  $\mu$. 

Our results for the higher derivatives $\chi_3^q/T$ and $\chi_4^q$ are shown in Fig.~\ref{highersusceptibilities}. Both observables peak in the 
crossover region near a pseudo-critical temperature. For $\chi_3^q/T$ and $\chi_4^q$ the error bars in the transition region are already
quite sizable due to the large fluctuations at the crossover. This is also the reason, why less values of $\mu \beta$ could be 
realiably evaluated. For the values we show, the 4th derivative displays only a rather small sensitivity to the chemical potential, which
is visible mainly in the crossover region where unfortunately also the error bars are largest. For large temperatures both, 
$\chi_3^q/T$ and $\chi_4^q$
approach the Stefan-Boltzmann values, and we observe that the relative deviations of the higher derivatives are smaller than for $n_q/T^3$ and $
\chi_2^q/T^2$.

From theoretical and experimental interest are ratios of the generalized susceptibilities. They are accessible in experiment and 
do not depend on the physical volume. In theory these quantities can give a clear signal of deconfinement with distinct 
behavior above, below and at the crossover. In addition they are essentially constant in the regions above and below the transition 
and one can compare them with model calculations like the hadron resonance gas (HRG) as stated in Section \ref{sec:hrg}. 
In Fig.~\ref{ratios} we show results for three different ratios of the quark number susceptibilities, $(\chi^q_2/T^2)/(n_q/T^3)$, 
$(\chi^q_3/T)/(\chi^q_2/T^2)$ and $(\chi^q_4)/(\chi^q_2/T^2)$ (top to bottom).
In addition to marking the Stefan-Boltzmann results (arrows on the right hand sides) we also show the results from the hadron resonance 
gas discussed in Section IV (dashed black lines in the confined region). 

In the high temperature region all ratios show very stable plateaus with small errors. For all values of $\mu \beta$ the positions of the plateaus 
are in almost perfect agreement with the corresponding Stefan-Boltzmann results (the ratio $(\chi^q_4)/(\chi^q_2/T^2)$ is 
independent of $\mu \beta$). Below the crossover the error bars are much larger, and only the ratio $(\chi^q_2/T^2)/(n_q/T^3)$ can be
studied in a quantitative way. However, for this case we find very good agreement with the model data, this time from the HRG. 

Concerning $(\chi^q_3/T)/(\chi^q_2/T^2)$ and $(\chi^q_4)/(\chi^q_2/T^2)$, at least for the 
$8^3 \times 4$ data where we have larger statistics, 
we find qualitative agreement with the HRG results, in particular the correct trend when 
increasing $\mu \beta$. 

The overall assessment is that the ratios are very accurately described by the Boltzmann limit above the crossover and 
at least for the lowest ratio we could establish good agreement with the HRG.  Between the two phases is a narrow region of transitory behavior 
connecting the two plateaus. 

\begin{figure}[t]
  \centering
  \hspace*{-5mm}
  \includegraphics[width=80mm,clip]{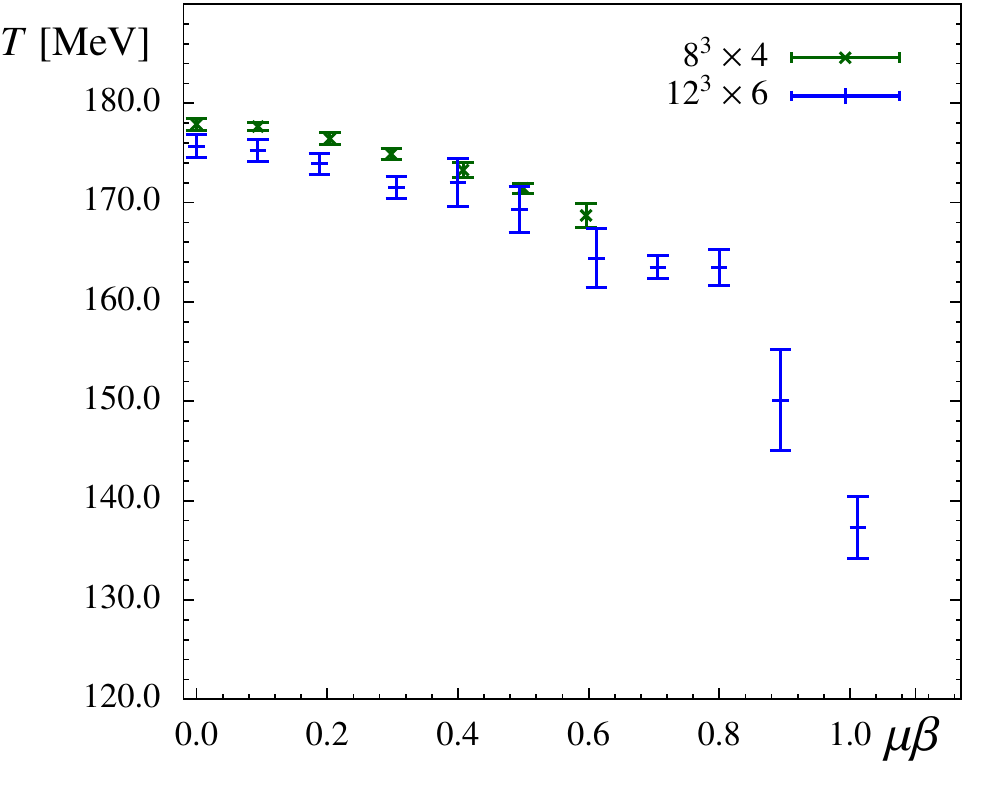}
  \caption{The critical line in the $\mu\beta$-$T$ plane, as determined from the inflection points of $\chi^q_2/T^2$.}
\label{phasediag}
\end{figure}

We conclude the presentation of our numerical results with an analysis of the critical line in the $\mu\beta$-$T$ plane. For that purpose 
we determined $T_c$ from the inflection point of our results for $\chi^q_2/T^2$ (compare the bottom plots in Fig.~4).  
In Fig.~\ref{phasediag} we show the corresponding data points in the $\mu\beta$-$T$ plane using results from our $8^3 \times 4$ and 
$12^3 \times 6$ lattices. The two data sets have a small discrepancy which, however, never exceeds
3 MeV. This is a surprisingly good agreement, since for the $8^3 \times 4$ one still expects sizable finite volume and 
discretization effects.  

\section{Summary} \label{sec:sum}

In this article we present results for generalized quark number susceptibilities and their ratios at finite temperature and chemical potential. The 
calculation is based on the fugacity expansion which we show to be particularly well suited for quark number related observables. In particular
we demonstrate that the susceptibilities can be obtained from moments of the expansion coefficients of the fugacity series, the so-called 
canonical determinants. The canonical determinants are needed very accurately for rather large values of the net quark number and we 
obtain them using Fourier transform with respect to imaginary chemical potential combined with a dimensional reduction technique. This 
numerical calculation is rather challenging, and part of the motivation for this article is to demonstrate, that the fugacity approach can indeed 
be extended to physically relevant volumes. 

Our calculations are done for two flavors of Wilson fermions on two lattice sizes with different resolution and temperature values on both sides
of the crossover.  We study the generalized susceptibilities up to fourth order and on the finer lattice could extract reliable results up to 
$\mu \beta \sim 1.0$ with moderate numerical effort. 
The results for the susceptibilities agree with other lattice studies and for the ratios we reproduce the 
expected good agreement with the HRG below the crossover and with the Stefan-Boltzmann results above. 

Although our results from the fugacity expansion are in good agreement with other calculations and the results from models,  
we are aware that the results presented here are still of exploratory nature. In particular a calculation on larger and finer lattices 
is needed which in turn would also allow one to use smaller quark masses.  Increasing the volume will drive up the cost since the range 
of the net quark numbers for which the canonical determinants  is needed is determined by the quark number susceptibility which is 
an extensive quantity. When lowering the quark mass studies in the free case suggest only a very weak increase in cost. 

To conclude, 
we believe that the fugacity expansion has an interesting potential as a complimentary expansion to the Taylor series, with different properties 
(infinite Taylor series versus finite Laurent series), and different numerical challenges.

\vskip3mm
\noindent
{\bf Acknowledgements: } The authors would like to thank Andrei Alexandru, Szabolcs Borsanyi,
Julia Danzer, Christian Lang, Bernd-Jochen Schaefer and Kim Splittorff, and Jac Verbaarschot 
for interesting discussions.  H.-P.~Schadler is funded by the FWF DK W1203, ``{\sl Hadrons in Vacuum, Nuclei and Stars}''. 
This work is partly supported by DFG TR55, ''{\sl Hadron Properties from Lattice QCD}'' and by the Austrian Science Fund FWF 
Grant. Nr. I 1452-N27.
\vskip8mm
 
\appendix*

\section{Dimensional reduction}\label{sec:app}
In this paper we use a modification of the formula for the dimensional reduction derived in \cite{dimreduction}. The modified form is more 
suitable for the applications discussed in this paper. In this appendix we use the conventions
and the notation of \cite{dimreduction}. Following  \cite{dimreduction}, after dividing the lattice into four domains which contain different timeslices, 
and integrating out the fermion degrees of freedom in all domains separately one obtains (this is Eq.~(10) in \cite{dimreduction})
\begin{equation}\label{eq:detDexp1}
  \det[D(\mu)]= \tilde A\, H(\mu)  \; .
\end{equation}
The factor $\tilde A$  is independent of the chemical potential and is given by a product of determinants of matrices
\begin{equation}
   \tilde A = \det[D^{(1)}]\det[D^{(3)}]\det[\tilde D^{(2)}]\det[\tilde D^{(4)}] \; ,
\end{equation}
where the $D^{(i)}$ denote the terms of the Dirac operator within the domains labeled by $i = 1,2,3,4$ and the $\tilde D^{(i)}$ are the $D^{(i)}$ 
plus some additional terms (see Eq.~(12) in \cite{dimreduction} for further details). The $\mu$-dependent part can be written as
\begin{equation}
	H(\mu) = \det[1-H_0 - e^{\mu \beta} H_+ - e^{-\mu \beta} H_-]  \; ,
	\label{aux3}
\end{equation}
where $H_0, H_+$ and $H_-$ are three different, $\mu$-independent, $N_s\times 1\times 3 \times 4$ dimensional matrices:
\begin{align*}
  H_0 =\, & \tilde S^{(4)}\,D^{(4,2)}_1\,\tilde S^{(2)}\,D^{(2,4)}_1 
        \\ &\quad+ \tilde S^{(4)}\,D^{(4,2)}_3\,\tilde S^{(2)}\,D^{(2,4)}_3 \; , \notag\\
  H_+ =\, & \tilde S^{(4)}\,D^{(4,2)}_1\,\tilde S^{(2)}\,D^{(2,4)}_3 \; , \notag\\
  H_- =\, & \tilde S^{(4)}\,D^{(4,2)}_3\,\tilde S^{(2)}\,D^{(2,4)}_1 \; ,
\end{align*}
with $\tilde S^{(i)} = (\tilde D^{(i)})^{-1}$, and the $D^{(i,j)}_k$ denote certain combinations of the Dirac terms connecting 
the domains and some $\tilde S^{(i)}$. 

Now we want to rewrite Eq.~(\ref{aux3}) such that the matrices 
attached to the factors $e^{\mu \beta}$ and $e^{-\mu \beta}$ are hermitian conjugate 
to each other, 
\begin{widetext}
\begin{align}
  \det[D(\mu)]&= \tilde A\, \det[1-H_0 - e^{\mu \beta} H_+ - e^{-\mu \beta} H_-]) \notag\\
    &= \tilde A\, \det[1-\tilde S^{(4)}\tilde H_0 - e^{\mu \beta} \tilde S^{(4)}\gamma_5 H_-^\dagger\gamma_5\tilde D^{(4)} - e^{-\mu \beta} H_-)] \notag\\ 
    &= \tilde A\, \det[\tilde S^{(4)}(\tilde D^{(4)}\gamma_5-\tilde H_0\gamma_5 - e^{\mu \beta} \gamma_5 H_-^\dagger\gamma_5\tilde D^{(4)}\gamma_5 - e^{-\mu \beta}\tilde D^{(4)} H_-\gamma_5)\gamma_5]  \notag\\ 
    &= \tilde A\det[\tilde S^{(4)}]\, \det[(\tilde D^{(4)}-\tilde H_0)\gamma_5 - e^{\mu \beta} \gamma_5 H_-^\dagger\tilde D^{(4)\dagger} - e^{-\mu \beta}\tilde D^{(4)} H_-\gamma_5]\det[\gamma_5] \; .
\end{align}
\end{widetext}
We have used the relations $\gamma_5\gamma_5=1$ and $\gamma_5\tilde D^{(4)}\gamma_5=\tilde D^{(4)\dagger}$ and a relation that
connects $H_+$ and $H_-$ (see \cite{dimreduction}). 
In the last step we have grouped the terms in such a way that we can identify two new matrices
\begin{align}
  &K_0 = (\tilde D^{(4)}-\tilde H_0)\gamma_5 \; ,\notag\\
  &K = \gamma_5 H_-^\dagger \tilde (D^{(4)})^\dagger = D^{(4,2)}_1\,\tilde S^{(2)}\,D^{(2,4)}_3\,\gamma_5\; ,
\end{align}
where 
\begin{equation}
\tilde H_0 = D^{(4,2)}_1\,\tilde S^{(2)}\,D^{(2,4)}_1+D^{(4,2)}_3\,\tilde S^{(2)}\,D^{(2,4)}_3 \; .
\end{equation}
We end up with the modified formula for the dimensional reduction of the fermion determinant (compare Eqs.~(\ref{dimred1}) and 
(\ref{dimred2}) in the main text),
\begin{equation}
  \det[D(\mu)]= A\, W(\mu \beta)  \; ,
\end{equation}
where the $\mu$-dependent factor is now given by
\begin{equation}
	W(\mu \beta) = \det[K_0 - e^{\mu \beta} K - e^{-\mu \beta} K^\dagger]  \; .
\end{equation}
The factor $A=\tilde A\det[\tilde S^{(4)}]$ is still independent of the chemical potential and drops out in calculations of observables (see Section 
\ref{sec:fug}). In contrast to three matrices in the original form of the winding expansion stated in \cite{dimreduction}, we now have just two distinct 
matrices $K_0$ and $K$. These matrices have to be pre-computed only once in the simulation
code and can then be used in the calculation for different 
values of chemical potential as described in the main text of the paper.


\vfill

\begin{thebibliography}{12} 
 
\bibitem{review}
  D.~Sexty, PoS LATTICE {\bf 2014} (2014) [arXiv:1410.8813];
%
  C.~Gattringer,
  PoS LATTICE {\bf 2013} (2013) 002  [arXiv:1401.7788];
%
  G.~Aarts,
  PoS LATTICE {\bf 2012} (2012) 017,
  [arXiv:1302.3028 [hep-lat]];
  L.~Levkova,
  PoS LATTICE {\bf 2011} (2011) 011,
  [arXiv:1201.1516 [hep-lat]].
 
\bibitem{oldfugacity}
  J.~Danzer, C.~Gattringer, 
  Phys. Rev. D {\bf 86} (2012) 014502, 
  [arXiv:1204.1020 [hep-lat]].

\bibitem{lattice2014}
H.-P.~Schadler and C.~Gattringer,
  PoS LATTICE {\bf 2014} (2014), 
  [arXiv:1409.4672 [hep-lat]].

\bibitem{modelfugacity}
  E.~Gr\"unwald, Y.~Delgado~Mercado, C.~Gattringer, 
  PoS LATTICE {\bf 2013} (2013) 448,
  [arXiv:1310.6520 [hep-lat]].
  E.~Gr\"unwald, Y.~Delgado~Mercado and C.~Gattringer,
  Int. Journal of Mod. Phys. A  Vol. 29, No. 32 (2014), [arXiv:1403.2086 [hep-lat]].
 
\bibitem{oldscale}
  A.~Alexandru, M.~Faber, I.~Horvath, K.-F.~Liu, 
  Phys. Rev. D {\bf 72} (2005) 114513, 
  [arXiv:hep-lat/0507020].
  
\bibitem{hasentouss}
A. Hasenfratz and D. Toussaint, Nucl. Phys. B 371 (1992) 539.

\bibitem{canonical}
  P.~de Forcrand and S.~Kratochvila,
  Nucl.\ Phys.\ Proc.\ Suppl.\  {\bf 153} (2006) 62
  [hep-lat/0602024];
%
  A.~Li, A.~Alexandru and K.~F.~Liu,
  Phys.\ Rev.\ D {\bf 84} (2011) 071503
  [arXiv:1103.3045 [hep-ph]].
 
\bibitem{fluctuations}
  B.~Friman, F.~Karsch, K.~Redlich, V.~Skokov, Eur. Phys. J. C {\bf 71} (2011) 1694, 
  [arXiv:1103.3511 [hep-ph]];  
%
  A. Bazavov {\it et al}, 
  Phys. Rev. D {\bf 86} (2012) 034509, [arXiv:1203.0784 [hep-lat]];
%
  A. Bazavov {\it et al}, 
  Phys.Rev. D88 (2013) 9, 094021, [arXiv:1309.2317 [hep-lat]];
%
  S.~Borsanyi, Z.~Fodor, S.~D.~Katz, S.~Krieg, C.~Ratti, K.~K.~Szabo, 
  Phys. Rev. Lett. {\bf 111} (2013) 062005,
  [arXiv:1305.5161 [hep-lat]];
%
  S.~Borsanyi, Z.~Fodor, S.~D.~Katz, S.~Krieg, C.~Ratti, K.~K.~Szabo,  
  Phys. Rev. Lett. {\bf 113} (2014) 052301, [arXiv:1403.4576 [hep-lat]].     

\bibitem{dimreduction}
  J.~Danzer, C.~Gattringer, 
  Phys. Rev. D {\bf 78} (2008) 114506, 
  [arXiv:0809.2736 [hep-lat]].
  
\bibitem{dimred2}
 K.~Nagata and A.~Nakamura,
  Phys.\ Rev.\ D {\bf 82} (2010) 094027
  [arXiv:1009.2149 [hep-lat]];
%
 A.~Alexandru and U.~Wenger,
  Phys.\ Rev.\ D {\bf 83} (2011) 034502
  [arXiv:1009.2197 [hep-lat]].
  
\bibitem{MILC}
  MILC collaboration,\\http://physics.utah.edu/\textasciitilde detar/milc.html

\bibitem{quarkdistribution}
  P.~Braun-Munzinger, B.~Friman, F.~Karsch, K.~Redlich, V.~Skokov, 
  Phys. Rev. C {\bf 84} (2011) 064911, 
  [arXiv:1107.4267 [hep-ph]]. 
  K.~Morita, B~Friman, K~Redlich, V~Skokov, 
  Phys. Rev. C {\bf 88} (2013) 034903, 
  [arXiv:1301.2873 [hep-ph]].

\bibitem{kim}
A.~Alexandru, C.~Gattringer, H.-P.~Schadler, K.~Splittorff and J.~J.~M.~Verbaarschot,
  arXiv:1411.4143 [hep-lat].
 
\bibitem{hrg}
  R.~Hagedorn, 
  CERN Report {\bf 71-72} (1971). 
  F.~Karsch, K.~Redlich, A.~Tawfik, 
  Eur. Phys. J. C {\bf 29} (2003) 549-556, 
  [arXiv:hep-ph/0303108]. 
  S.~Ejiri, F.~Karsch, K.~Redlich, 
  Phys. Lett. B {\bf 633} (2006) 275-282, 
  [arXiv:hep-ph/0509051]. 
  A.~Andronic, P.~Braun-Munziger, J.~Stachel, M.~Winn, 
  Phys. Lett. B {\bf 718} (2012) 80-85, 
  [arxiv:1201.0693 [nucl-th]].

\bibitem{nana}
  A.~Nakamura and K.~Nagata,
  arXiv:1305.0760 [hep-ph].
%
  A.~Nakamura and K.~Nagata,
  Nucl.\ Phys.\ A {\bf 931} (2014) 825.
  


\end{thebibliography}
\end{document}